\documentclass[11pt]{article}
\usepackage{amsmath,amssymb,amsfonts,epsf}
\def\ci{\mathrm{i}}
\def\ud{\mathrm{d}}
\def\Tr{\mathrm{Tr}}

\newcommand{\expp}{\mathrm{e}}
\renewcommand{\Re}{\operatorname{Re}}

\renewcommand{\v}{\boldsymbol v}
\newcommand{\s}{\boldsymbol s}
\renewcommand{\u}{\boldsymbol u}
\newcommand{\q}{\boldsymbol q}
\newcommand{\p}{\boldsymbol p}

\begin{document}
~\vspace{3.0cm}

\centerline{\LARGE The semiclassical relation between open trajectories}
\vspace{0.3cm}
\centerline{\LARGE and periodic orbits for the Wigner time delay}
\vspace{2.0cm}
\centerline{\large 
Jack Kuipers\footnote{E-mail: jack.kuipers@bristol.ac.uk}
and Martin Sieber\footnote{E-mail: m.sieber@bristol.ac.uk}}
\vspace{0.5cm}

\centerline{School of Mathematics, University of Bristol,
Bristol BS8\,1TW, UK}

\vspace{4.0cm}
\centerline{\bf Abstract}
\vspace{0.5cm}
The Wigner time delay of a classically chaotic quantum system can be expressed 
semiclassically either in terms of pairs of scattering trajectories that
enter and leave the system or in terms of the periodic orbits trapped
inside the system.  We show how these two pictures are related on the
semiclassical level. We start from the semiclassical formula with the
scattering trajectories and derive from it all terms in the periodic
orbit formula for the time delay. The main ingredient in this calculation
is a new type of correlation between scattering trajectories which is due
to trajectories that approach the trapped periodic orbits closely.
The equivalence between the two pictures is also demonstrated by considering
correlation functions of the time delay. A corresponding calculation for
the conductance gives no periodic orbit contributions in leading order.

\vspace{2.5cm}

\noindent PACS numbers: \\
\noindent 03.65.Sq ~ Semiclassical theories and applications. \\
\noindent 05.45.Mt ~ Semiclassical chaos (``quantum chaos'').

\clearpage

\section{Introduction}

Quantum systems whose classical counterparts are chaotic show universal 
statistical fluctuations that are well modelled by Random Matrix Theory 
(RMT) \cite{bohigas84,haake01}. In the semiclassical limit, formally as
$\hbar \to 0$, the quantum statistics can be approximated by quantities
that involve trajectories of the classical dynamics. One focus of work
in semiclassics has been to recreate results from RMT by semiclassical
methods.

For closed systems, statistical fluctuations in the energy spectrum can
be seen, for example, in the spectral form factor $K(\tau)$ which is
semiclassically approximated by a double sum over periodic orbits. The
universal fluctuations from RMT are then due to correlated pairs of
periodic orbits. The diagonal approximation, comparing an orbit to itself
(or its time reverse) leads to the first term in the expansion of the form
factor \cite{ha84, berry85}. All the other terms in the small time ($\tau<1$)
expansion of the form factor can be obtained, in agreement with RMT, from
periodic orbits with self-encounters. These are events where an orbit
approaches itself (or its time reverse) very closely so that its partner
can cross the encounter region differently \cite{sr01,muller04}.
Currently it is not known what other type of periodic orbit correlations
contribute in the regime $\tau>1$, though their contribution is calculated
indirectly in \cite{heusler07,KM07}. Similar methods as for the form factor
have been applied, for example, to the conductance through open systems
where the semiclassical sum is over pairs of open trajectories that start
and end in the leads \cite{rs02,heusler06,muller07}.

For the Wigner time delay in open systems we have the interesting situation
that the semiclassical approximation can be expressed in two ways. One
gives the time delay as a double sum over scattering trajectories that
enter and leave the system, in a similar way as for the conductance. The other is
through a relation to a density of states and leads to a semiclassical
formula that contains the average time delay plus a {\em single}
sum over the periodic orbits that are trapped in the system. Our main
motivation for this article is to understand the duality of these two
semiclassical pictures. For open chaotic cavities, 
we will start from one of these pictures, the
double sum over scattering trajectories, and derive all terms in the
other semiclassical formula for the time delay. The periodic orbit
terms are obtained by considering a new type of open trajectory correlation
which is linked to the motion around periodic orbits. For systems
without time-reversal symmetry these {\em periodic orbit encounters}
are sufficient for obtaining the correct periodic orbit terms. For
systems with time-reversal symmetry one also has to include combinations
of self-encounters and periodic orbit encounters.

We will also consider a correlation function of the time delay. When expressed
as a double sum over periodic orbits, the diagonal approximation
\cite{eckhardt93, val98} and higher order terms \cite{ks07} were shown
to agree with RMT (for a small time expansion). Using the other semiclassical
approximation which is in terms of a quadruple sum over open trajectories,
it was shown in \cite{lv04} that the diagonal approximation does
not give the leading RMT result. This was attributed to the non-unitarity
of the semiclassical scattering matrix \cite{vl01}. We will show that
the inclusion of off-diagonal terms due to trajectories with self-encounters
removes the discrepancy with RMT and restores the semiclassical unitarity
of the scattering matrix. These calculations are similar to those
for correlation functions of the conductance \cite{braun06, muller07}.

Our paper is divided as follows.  In section~\ref{timedelay} we introduce
the two semiclassical approaches for the time delay. In section~\ref{average}
we show which correlated pairs of open trajectories recreate the average time
delay.  We then go beyond the average to recreate the periodic orbit contributions
in section~\ref{periodic}. This is achieved by introducing the new type of 
trajectory correlations which are due to periodic orbit encounters. The correlation
function of the time delay is considered in section~\ref{correl}, and our
conclusions follow in section~\ref{concl}.

\section{The time delay} \label{timedelay}

For a chaotic cavity with one or more open leads that carry $M$ scattering
channels, the incoming and outgoing waves are related by the $M \times M$
scattering matrix $S(E)$.  The Wigner time delay, which represents the extra
time spent in the scattering process compared to free motion, is defined as 
\cite{wigner55, smith60}
\begin{equation} \label{timedelayeqn}
\tau_{\mathrm{W}}(E) = -\frac{\ci\hbar}{M}
\Tr \left[ S^{\dagger}(E) \frac{\ud S(E)}{\ud E} \right] 
= - \frac{\ci\hbar}{M} \frac{\ud}{\ud E} \ln \det S(E) \; .
\end{equation}
The time delay can be expressed semiclassically both in terms of the 
trapped set of periodic orbits of the open system, and in terms of the 
open scattering trajectories that enter and exit through the leads.  

The description of the Wigner time delay in terms of trapped periodic orbits 
comes from its relation to a density of states which, in general, is the 
difference between the level density of the open scattering system and a
free system \cite{friedel52}
\begin{equation} \tau_{\mathrm{W}}(E) = \frac{2\pi\hbar}{M}d(E)
\approx\frac{2\pi\hbar}{M} \bar{d}(E) + \frac{2\pi\hbar}{M} d^{\mathrm{fl}}(E) \; ,
\end{equation}
Here the density of states $d(E)$ is separated into a mean part 
$\bar{d}(E)$ and a fluctuating part $d^{\mathrm{fl}}(E)$ which each have 
a semiclassical approximation.  The approximation for the mean density of 
states, for a chaotic cavity with $2$ degrees of freedom, comes from Weyl's law 
for the corresponding closed system $\bar{d}(E)\sim\Omega/(2\pi\hbar)^{2}$ 
where $\Omega$ is the phase space volume of the shell of constant energy $E$.
The fluctuating part can be expressed, like in the Gutzwiller trace formula 
\cite{gutzwiller71} as a sum over the periodic orbits.  The difference is 
that the sum only includes periodic orbits that are trapped in the system 
\cite{bb74, val98}.  Using these approximations, we can write the time delay as
\begin{equation} \label{timedelayeq}
\tau_{\mathrm{W}}(E) \approx \frac{T_{\mathrm{H}}}{M} + \frac{2}{M} \Re
\sum_{p,r} A_{p,r}(E) \expp^{\frac{\ci}{\hbar} r S_{p}(E)}
\expp^{-\frac{\ci\pi}{2} r \mu_{p}} \; ,
\end{equation}
where $T_{\mathrm{H}}$ is the Heisenberg time which is related to the 
average level density by $T_{\mathrm{H}}=2\pi\hbar\bar{d}(E)$. The first term 
is the average time spent in the cavity $\bar{\tau}_{\mathrm{W}}=T_{\mathrm{H}}/M$.
It is equal to the inverse of the classical escape rate which can be expressed
in the form $\mu = M/T_{\mathrm{H}}$ \cite{lv04}. In 
the sum $p$ labels the trapped primitive periodic orbits and $r$ 
their repetitions.  The orbits have action $S_{p}$ and Maslov index
$\mu_{p}$.  Their stability amplitude $A_{p,r}$ can be written
in terms of the stability matrix $M_{p}$ and the period $T_{p}$
\begin{equation} \label{poamp}
A_{p,r}= \frac{T_p}{\sqrt{\vert \det(M_{p}^{r}-1)\vert}} \; .
\end{equation}
The description of the time delay in terms of open trajectories comes 
from the semiclassical approximation to the scattering matrix elements 
\cite{miller75,richter00,rs02}
\begin{equation} \label{scatmateqn}
S_{ba}(E) \approx \frac{1}{\sqrt{T_{\mathrm{H}}}}\sum_{\alpha (a \to b)}
A_{\alpha}\expp^{\frac{\ci}{\hbar}S_{\alpha}}
\expp^{-\frac{\ci\pi}{2}\nu_{\alpha}} \; .
\end{equation}
Here $S_{\alpha}$ is the action of the trajectory $\alpha$ and
$\nu_{\alpha}$ is the number of conjugate points along the trajectory
(plus twice the number of reflections on walls with Dirichlet boundary
conditions). The stability amplitude $A_\alpha$ can be found in \cite{richter00}.
The sum is then over all classical trajectories that start in channel $a$ and end 
in channel $b$, where the channels fix the absolute value of the angles at
which the trajectories enter and leave the cavity.  From this semiclassical
approximation for the scattering 
matrix elements we can obtain an expression for the time delay by 
substituting into equation~(\ref{timedelayeqn}).  When we differentiate 
the scattering matrix elements we ignore the change in the 
slowly varying prefactor and only keep the term from the oscillating action 
exponentials
\begin{equation} \label{timedelaytraj}
\tau_{\mathrm{W}}\approx\frac{1}{MT_{\mathrm{H}}}\sum_{a,b}
\sum_{\alpha, \alpha'(a\to b)}T_{\alpha}A_{\alpha}A^{*}_{\alpha'}
\expp^{\frac{\ci}{\hbar}(S_{\alpha}-S_{\alpha'})}
\expp^{-\frac{\ci\pi}{2}(\nu_{\alpha}-\nu_{\alpha'})} \; ,
\end{equation}
where $T_{\alpha}=\partial{S_{\alpha}}/\partial{E}$ is the time the 
trajectory $\alpha$ spends inside the system.  Here we can see that the 
time delay is a sum over trajectory pairs $\alpha,\alpha'$ both of which start 
and end in the same channels ($a$ and $b$ respectively), followed by a sum 
over all the possible channels.

We can also consider a correlation function of scattering matrix elements 
\begin{equation}
C(\epsilon)=\sum_{a,b}S_{ba}\left(E+\frac{\epsilon M }{4 \pi \bar{d}}\right)
S_{ba}^{*}\left(E-\frac{\epsilon M }{4 \pi \bar{d}}\right) \; ,
\end{equation}
where it is convenient to specify the energy difference in units of
$M (2 \pi \bar{d})^{-1} = \hbar \mu$, because this will simplify the formulae
in the following. If we set $\epsilon=0$ this becomes
\begin{equation}
C(0)=\Tr\left[S(E) S^{\dagger}(E)\right] \; .
\end{equation}
By using the semiclassical approximation of the matrix elements from 
equation~(\ref{scatmateqn}) and expanding the action up to first order 
in energy, $S_{\alpha}(E+\eta)\approx S_{\alpha}(E)+\eta T_{\alpha}(E)$, 
the correlation function can be expressed in terms of pairs of 
scattering trajectories
\begin{equation} \label{cepssemi}
C(\epsilon)\approx\frac{1}{T_{\mathrm{H}}}\sum_{a,b}
\sum_{\alpha, \alpha'(a\to b)}A_{\alpha}A^{*}_{\alpha'}
\expp^{\frac{\ci}{\hbar}(S_{\alpha}-S_{\alpha'})}
\expp^{-\frac{\ci\pi}{2}(\nu_{\alpha}-\nu_{\alpha'})}
\expp^{\frac{\ci \epsilon \mu}{2} (T_{\alpha}+T_{\alpha'})} \; ,
\end{equation}
from which we can obtain a symmetrized version of the time delay \cite{lv04} 
\begin{equation} \label{timedelaytrajsym}
\tau_{\mathrm{W}} = \frac{-\ci}{\mu M}\frac{\ud}{\ud \epsilon}
C(\epsilon)\Big\vert_{\epsilon=0} = -\frac{\ci\hbar}{2 M}
\Tr \left[ S^{\dagger}(E) \frac{\ud S(E)}{\ud E}  -
           S(E) \frac{\ud S^{\dagger}(E)}{\ud E}  \right] \; .
\end{equation}
Equation (\ref{timedelaytrajsym}) agrees with the definition of the time delay
in (\ref{timedelayeqn}) because of the unitarity of the scattering matrix.
If we insert (\ref{cepssemi}) into (\ref{timedelaytrajsym}) we obtain again
a semiclassical formula for the time delay which differs slightly from
(\ref{timedelaytraj}) in that the time $T_\alpha$ is replaced by the
average time $(T_{\alpha}+T_{\alpha'})/2$. Both formulae are equivalent
and the difference only plays a role in section~\ref{correl} where it will be
discussed. We use the relation of the time delay to the function
$C(\epsilon)$ in the following to simplify the calculation by exploiting its
similarity to the average conductance of a chaotic ballistic device
\cite{heusler06,muller07}. In particular we will obtain, as in the case of the conductance,
simple diagrammatic rules for the semiclassical contributions of correlated
trajectories.

Because we can express the time delay equally in terms of open trajectories 
and trapped periodic orbits, there should be a semiclassical equivalence 
between the two pictures and the following should hold
\begin{equation} \label{timedelayequiv}
\frac{1}{MT_{\mathrm{H}}}\sum_{a,b}\sum_{\alpha, \alpha'(a\to b)}
T_{\alpha}A_{\alpha}A^{*}_{\alpha'}\expp^{\frac{\ci}{\hbar}(S_{\alpha}-S_{\alpha'})}
\expp^{-\frac{\ci\pi}{2}(\nu_{\alpha}-\nu_{\alpha'})}
\approx \bar{\tau}_{\mathrm{W}}+\frac{2}{M} \Re \sum_{p,r}
A_{p,r}(E)\expp^{\frac{\ci}{\hbar} r S_{p}(E)}
\expp^{-\frac{\ci\pi}{2}r \mu_{p}} \; .
\end{equation}
The left hand side is the sum over scattering trajectories, while the right 
includes an average part and a sum over trapped periodic orbits.  We shall 
now show how, by considering the contributions in the semiclassical limit of 
pairs of correlated trajectories from the sum on the left we can recreate 
all the terms on the right. We will consider only systems with two degrees
of freedom, but the calculations are very similar in higher dimensions
\cite{muller04}. The assumptions for the semiclassical calculations in this
article are the same as for the conductance in \cite{muller07}.
In particular, corrections related to the finite value of the quotient
of the Ehrenfest time and the dwell time are neglected.

\section{The average time delay} \label{average}

In this section we shall derive the average time delay from correlated
pairs of open trajectories. For systems without time-reversal symmetry
the diagonal approximation (pairing a trajectory with itself) suffices
\cite{lv04}.  For systems with time-reversal symmetry, however, a small
correction is needed which comes from trajectories that have close
self-encounters.  The calculation follows similar steps and ideas
to the calculation of the average conductance \cite{rs02,heusler06,muller07},
which in turn builds on work on spectral statistics \cite{muller04}.
We exploit this similarity by concentrating on the correlation function
of the scattering matrix elements $C(\epsilon)$.
Its semiclassical approximation is given by (\ref{cepssemi}) which, 
aside from the exponential factor containing the trajectory time $T_{\alpha}$
and a different channel sum, is the same as the conductance from 
\cite{muller07}, so we only highlight the relevant points in the calculation.
We shall be more detailed in section \ref{periodic} where we
introduce the new correlations that produce the periodic orbit terms. 

The diagonal term for the sum over trajectories that connect channels $a$
and $b$ considers pairs where the two trajectories 
$\alpha$ and $\alpha'$ are identical, and it gives a contribution of
\begin{equation} \label{diag1}
\frac{1}{T_{\mathrm{H}}} \sum_{\alpha (a \to b)} \vert A_{\alpha}\vert^{2}
\expp^{\ci \epsilon \mu T_{\alpha}} \; .
\end{equation}
The sum in (\ref{diag1}) can be performed by using a sum rule for open
trajectories \cite{rs02} which turns it into an integral over the
trajectory time $T$
\begin{equation} \label{opensumrule}
\sum_{\alpha(a\to b)}\vert A_{\alpha}\vert^{2} \ldots
\approx \int_{0}^{\infty}\ud T \: \expp^{-\mu T} \ldots \; ,
\end{equation}
The exponential term in (\ref{opensumrule}) represents the
average probability that a trajectory
remains in the system  for the time $T$.  The sum over channels depends on the symmetry 
of the dynamics.  For systems without time-reversal symmetry ($\kappa=1$) 
we can pick both $a$ and $b$ from the $M$ possible channels giving a factor 
of $M^{2}$.   For systems with time-reversal symmetry ($\kappa=2$) we can
also pair the trajectory $\alpha$ with its time reversal if the
start and end channel are the same ($a=b$), and this contributes an
additional $M$ to the channel sum.  The diagonal approximation thus becomes
\begin{equation}
C^{\mathrm{diag}}(\epsilon)\approx\frac{M(M+\kappa-1)}{T_{\mathrm{H}}}
\int_{0}^{\infty}\ud T \: \expp^{-\mu (1 - \ci\epsilon)T}
=\frac{M(M+\kappa-1)}{M(1 - \ci \epsilon)} \; .
\end{equation}
\begin{figure}
\begin{center}
\mbox{\epsfxsize6cm\epsfbox{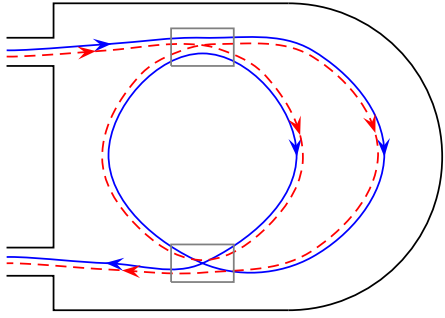}}
\end{center}
\caption{An example of a trajectory (full line) with two encounter regions
and its partner trajectory (dashed line). The encounter regions are
indicated by rectangular boxes. \label{trajpic}}
\end{figure}
Using (\ref{timedelaytrajsym}) this already leads to the correct result
for the average time delay for systems without time reversal symmetry,
but not for systems with time-reversal symmetry (because the prefactor
then contains $M(M+1)$ instead of $M^2$). Hence we need to consider
off-diagonal contributions to the average of $C(\epsilon)$. They come from long 
trajectories that have one or more self-encounters in which two or more
stretches of a trajectory are almost identical. In systems with time-reversal
symmetry the stretches can also be almost mutually time-reversed.
The encounter regions are connected to each other and to the entry
and exit channels by long parts of the trajectory called {\em links}.
An example of a trajectory with two encounter regions is shown in
figure \ref{trajpic}. The partner trajectory connects the links in
a different way in the encounter regions, but follows the original
trajectory very closely along the links. Both encounter regions in
the figure are examples of 2-encounters which are traversed by two
stretches of an orbit. In general an arbitrary number of $l \geq 2$
stretches of an orbit can be almost identical (up to time-reversal)
in an encounter region and one then speaks of an $l$-encounter.
The total numbers of the different encounter regions of a trajectory
are collected in a vector $\v$ whose components $v_l$ specify
the numbers of $l$-encounters of a trajectory. For the calculation
of the off-diagonal terms one has to consider all possible {\em structures}
or {\em families}\footnote{The authors of \cite{muller07} use the
expression {\em structures} in the context of periodic orbits and
{\em families} in the context of open trajectories},
i.e.\ all topologically distinct ways in which two trajectories can
be correlated. A more precise definition of {\em families} can be found
in \cite{muller07}, but we shall need in the following only the fact
that there is a sum rule for the number $N(\v)$ of
different {\em families} with the same vector $\v$. 
Further relevant quantities are the total number of encounters of a
trajectory $V = \sum_l v_l$, and the total number of orbit stretches
of a trajectory in the encounter regions $L = \sum_l l \, v_l$.
The total number of links of a trajectory is then $L+1$. 

The action difference of the trajectories is given in terms of coordinates
along the stable and unstable manifolds in Poincar\'e surfaces of sections
in the $V$ encounter regions. These coordinates describe the relative positions of the
different stretches of a trajectory. The encounter regions are labeled by 
$\sigma$ and the number of trajectory stretches in them by $l_\sigma$. 
In the linearized approximation the action difference is given by
\begin{equation}
S_{\alpha}-S_{\alpha'} \approx \sum_{\sigma=1}^{V}\s_{\sigma}\u_{\sigma}=\s\u \; ,
\end{equation}
where $\s_\sigma$ and $\u_\sigma$ are vectors with dimension $(l_\sigma-1)$,
and $\s$ and $\u$ contain the components of all these vectors.

The summation over the correlated trajectory pairs is simplified by the fact
that different {\em families} with the same vector $\v$ give the same contribution
\cite{muller07}. Hence it is convenient to collect all these contributions and
sum over all trajectories pairs whose correlations are specified by the vector $\v$.
As for the diagonal approximation, we have also to take into account that one
can pair a trajectory with the time-reverse of its partner orbit if $a=b$ in
systems with time-reversal symmetry, and this gives an additional factor of two
in these cases. This factor is denoted by $\eta_{ab}$ in the following 
\begin{equation} \label{cv1}
C^{\v}(\epsilon)\approx\frac{1}{T_{\mathrm{H}}}\sum_{a,b} \eta_{ab}
\sum_{\alpha,\alpha' (a\to b)}^{\mathrm{fixed} \, \v}
|A_{\alpha}|^2 \expp^{\frac{\ci}{\hbar}(S_{\alpha}-S_{\alpha'})}
\expp^{\ci \epsilon \mu T_{\alpha}} \; ,
\end{equation}
where the approximations $T_{\alpha} \approx T_{\alpha'}$,
$A_{\alpha} \approx A_{\alpha'}$ and
$\mu_{\alpha} \approx \mu_{\alpha'}$ have been made, and
\begin{equation} \label{etadef}
\eta_{ab} = 1 + (\kappa - 1) \delta_{ab} \; .
\end{equation}

The sum over the trajectory pairs in (\ref{cv1}) is performed
by applying an ergodicity argument together with the finite escape
probability of the scattering trajectories which results in replacing
it by an integral
\begin{equation} \label{repl}
\sum_{\alpha,\alpha' (a\to b)}^{\mathrm{fixed} \, \v} |A_{\alpha}|^2 \ldots
\approx N(\v) \int \ud T \int \ud \s \, \ud \u \; w_{\v,T}(\s,\u) \, 
\expp^{-\mu T_{\mathrm{exp}}} \ldots \; .
\end{equation}
Here $N(\v)$ is the number of {\em families} with the same vector $\v$,
and $T$ is the time of the trajectories. $T_{\mathrm{exp}}$ will be
specified below, and $w_{\v,T}(\s,\u)$ is the probability density that
a trajectory of time $T$ has self-encounters specified by the separation
coordinates $\s$ and $\u$. It is given explicitly in terms of an integral
over $L$ of the $L+1$ link times
\begin{equation} \label{wvt}
w_{\v,T}(\s,\u)=\int' \ud t_1 \ldots \ud t_L \; \frac{1}{\Omega^{L-V}
\prod_{\sigma=1}^{V} t_{\mathrm{enc}}^{\sigma}(\s,\u)} \; ,
\end{equation}
where $t_{i}$ is the time of link $i$ and $t_{\mathrm{enc}}^{\sigma}$ is the 
time of encounter in the encounter region $\sigma$. The prime at the integral
denotes that it is subject to the restrictions that the link times must be
positive. Furthermore, the total
time of the links and encounter stretches is the time of the trajectory
\begin{equation} \label{tt}
T = \sum_{i=1}^{L+1}t_{i}+\sum_{\sigma=1}^{V}l_{\sigma}t_{\mathrm{enc}}^{\sigma} \; ,
\end{equation}
and the time of the last link $t_{L+1}$, which is fixed by $T$ and the $L$ other
link times, has to be positive too. The encounter times $t_{\mathrm{enc}}^\sigma$
are specified by requiring that all the components of $\s_\sigma$ and $\u_\sigma$
that determine the separation of the different stretches of a trajectory in the
encounter region $\sigma$ have a modulus that is smaller than a small arbitrary
constant $c$, and it is given by
\begin{equation} \label{enctime1}
t_{\mathrm{enc}}^\sigma \approx \frac{1}{\lambda} \ln \frac{c^2}{\max_i |s_{\sigma_i}|
\times \max_j |u_{\sigma_j}|} \, ,
\end{equation}
where $\lambda$ is the Lyapunov exponent. The relevant encounter times are of the
order of the Ehrenfest time.
Finally, the {\em exposure} time $T_{\mathrm{exp}}$ differs slightly from the
time $T$ of the trajectory, because the trajectory stretches during each encounter
are very close together. If the trajectory survives during one crossing of the
encounter region it will
survive all the others crossings.  The exposure time, the effective time where the 
trajectory can leave, is thus given by 
$T_{\mathrm{exp}}=T - \sum_{\sigma} (l_\sigma - 1) t_{\mathrm{enc}}^{\sigma}$.
Putting everything together, i.e. inserting (\ref{repl}), (\ref{wvt}) and (\ref{tt})
into (\ref{cv1}), one obtains after a change of the integration variable $T$
to the last link time $t_{L+1}$ an expression that contains integrals over all
link times
\begin{equation} \label{Coffeqn}
C^{\v}(\epsilon) \approx \frac{N(\v)}{T_{\mathrm{H}}} 
\sum_{a,b} \eta_{ab} \prod_{i=1}^{L+1} \left( \int_{0}^{\infty} \ud t_{i} \,
\expp^{-\mu(1 - \ci \epsilon) t_{i}} \right) 
\prod_{\sigma=1}^{V} \left(\int \ud \s_\sigma \, \ud \u_\sigma 
\frac{\expp^{-\mu(1 - \ci\epsilon l_{\sigma}) t_{\mathrm{enc}}^{\sigma}}
\expp^{\frac{\ci}{\hbar} \s_\sigma \u_\sigma}}{\Omega^{l_{\sigma}-1}
t_{\mathrm{enc}}^{\sigma}} \right) \; .
\end{equation}
One can see that (\ref{Coffeqn}) factors into a product over the $L+1$ links and the $V$
encounter regions. This is the main advantage of working with the correlation
function $C(\epsilon)$ instead of directly with the time delay, because the corresponding
expression for the time delay does not factorize (because of the $T_\alpha$ in
the pre-exponential factor in (\ref{timedelaytraj})). The factorization property
will be useful also in the following sections. The integrals over coordinates
in the encounter regions can be performed by using the semiclassical result
\cite{muller04,muller05b}
\begin{equation} \label{suinteqn}
\int\ud\s_{\sigma} \, \ud\u_{\sigma} \; (t_{\mathrm{enc}}^{\sigma})^{k}
\expp^{\frac{\ci}{\hbar} a \s_{\sigma}\u_{\sigma}} 
\approx \begin{cases} 0 & \text{if} \; k=-1 \; , \\
\frac{1}{|a|} (2 \pi \hbar)^{(l_\sigma-1)} & \text{if} \; k=0 \; , \end{cases}
\end{equation}
where $a$ is a real constant. The asymptotics of these integrals comes from
the origin, or zero separation, at which point the semiclassical approximations
made above become accurate. We can now perform all the integrals in
equation~(\ref{Coffeqn}). The integral over the encounter regions is evaluated
by expanding the exponential that contains the encounter times up to the linear
term. Higher order terms in the expansion are neglected, because they are of the
order of the quotient of the Ehrenfest time and the dwell time. For every encounter
we obtain thus a contribution of $-\mu (1 - \ci \epsilon l_\sigma)
T_{\mathrm{H}}^{l_\sigma-1}$ where $\mu=M/T_{\mathrm{H}}$, and for every link a
contribution of $\mu^{-1} (1 - \ci \epsilon)^{-1}$. All the Heisenberg
times cancel, and we obtain the simple diagrammatic rules that every
link contributes a factor of $M^{-1} (1 - \ci \epsilon)^{-1}$, and every
encounter a factor of $-M (1 - \ci \epsilon l_\sigma)$. Altogether the result is
\begin{equation}
C^{\v}(\epsilon) \approx N(\v) \sum_{a,b} \eta_{ab}
\frac{(-1)^V \prod_{\sigma=1}^{V} (1 - \ci \epsilon l_{\sigma})}
{M^{L-V+1} (1 - \ci \epsilon)^{L+1}} \; .
\end{equation}
As for the diagonal approximation the sum over the channels $a$ and $b$ gives
a factor $M (M+\kappa-1)$. It is convenient to introduce $k=L-V+1$ and to
combine the contributions from all vectors $\v$ with the same value of $k$.
We can also include the diagonal term by saying it corresponds to a trajectory
with no encounter $L=V=0$ and $N(\v)=1$.  With the diagonal term, all these
trajectories give the average contribution to the correlation function 
\begin{equation}
\bar{C}(\epsilon)\approx M(M+\kappa-1)\sum_{k=1}^{\infty}
\sum_{\v}^{L-V+1=k}(-1)^{V}N(\v)\frac{\prod_{\sigma=1}^{V}
(1 - \ci \epsilon l_{\sigma})}{M^k (1 - \ci \epsilon)^{L+1}} \; .
\end{equation}
We expand the result around $\epsilon=0$ and use $\sum_{\sigma}l_{\sigma}=L$
to obtain
\begin{equation}
\bar{C}(\epsilon)\approx M(M+\kappa-1)\sum_{k=1}^{\infty}
\sum_{\v}^{L-V+1=k} \frac{(-1)^{V} N(\v)}{M^k}
[1 + \ci \epsilon + O(\epsilon^2) ] \; .
\end{equation}
The sum over the vectors $\v$ can now be evaluated by using a sum
rule for the numbers $N(\v)$ that was obtained from a recursion relation
in \cite{muller07}
\begin{equation} \label{recursion}
\sum_{\v}^{L-V+1=k}(-1)^{V} N(\v)=(1-\kappa)^{k-1} \; .
\end{equation}
This leaves the evaluation of a geometric series to obtain
\begin{equation} \label{cave}
\bar{C}(\epsilon) \approx
M [1 + \ci \epsilon + O(\epsilon^2) ] \; ,
\end{equation}
which is the final result. Equation (\ref{cave}) shows first that the
semiclassical approximation is consistent with the unitarity of the
scattering matrix, because $\bar{C}(0) =
\Tr S(E) S^\dagger(E) \approx M$. Secondly, one obtains
the correct average time delay from the term that is linear in $\epsilon$,
$\bar{\tau}_{\mathrm{W}} = T_{\mathrm{H}}/M$.

\section{Periodic orbit terms} \label{periodic}

\subsection{Periodic orbit encounters} \label{subsect1}

\begin{figure}
\begin{center}
\mbox{\epsfxsize10cm\epsfbox{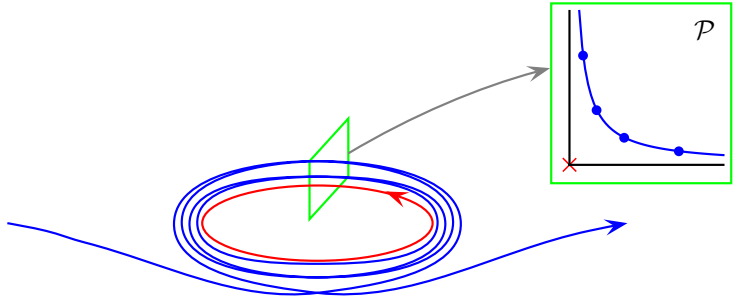}}
\end{center}
\caption{A schematic picture of a trajectory that approaches an unstable
periodic orbit follows it a number of times and leaves it again. In a
Poincar\'e map transverse to the periodic orbit the trajectory moves
along the invariant hyperbola $s u = \text{const}$. \label{orbpic}}
\end{figure}

The types of correlated trajectory pairs we will consider now approach
a trapped unstable periodic orbit, follow it very closely, and leave it again.
An example of such a trajectory is shown 
in figure~\ref{orbpic}. The trajectory approaches 
the periodic orbit almost along its stable manifold, goes around
a number of times, and then leaves it closely following the unstable 
manifold. In the vicinity of the periodic orbit we describe the motion
of the trajectory in a Poincar\'e surface of section transverse to the
periodic orbit. Because of the Birkhoff-Moser theorem we can make a
symplectic transformation to normal form coordinates which are along
the stable and unstable manifolds. Close to a periodic orbit one can
use the linearized approximation in which the Poincar\'e map has the
simple form \cite{almeida88}
\begin{equation} \label{map}
s' =\Lambda_p^{-1} s \, , \qquad u' =\Lambda_p u \, ,
\end{equation}
so that the trajectory moves along the invariant hyperbola $s u = \text{const}$.
$\Lambda_p$ and $1/\Lambda_p$ are the eigenvalues of the
stability matrix of the primitive periodic orbit $p$, with $|\Lambda_p|>1$.
If $\Lambda_p$ is negative the map involves also a reflection about
the origin. The correlated trajectories that we will consider differ
in the number of times they wind around the periodic orbit, and we shall
see that the semiclassical periodic orbit contribution is obtained
in the limit that $s u$ of these orbits goes to zero. This allows the
application of the linearized approximation.

We have to specify how to count the number of times a trajectory winds 
around a periodic orbit. This can be done by fixing an arbitrary small
positive constant $c$. The encounter region of the trajectory with the
periodic orbit is then defined by requiring that the moduli of the
coordinates $s$ and $u$ are both smaller than $c$. This is very similar
to the definition of the self-encounter regions in the previous section.
A very long trajectory has a finite probability to enter such an
encounter region.

\begin{figure}
\begin{center}
\mbox{\epsfxsize6cm\epsfbox{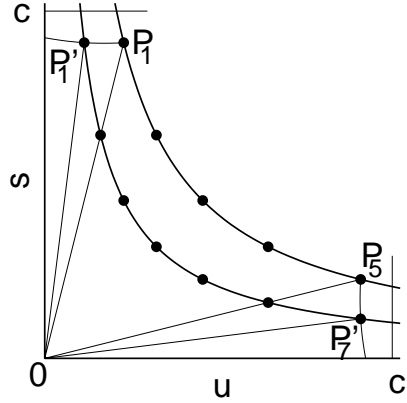}}
\end{center}
\caption{Two trajectories which follow the periodic orbit at $O$ five and seven times,
respectively, within the region of the Poincar\'e section bounded by the constant $c$.
The action difference is $S_{\alpha}-S_{\alpha'}=A - 2 S_p$ where $S_p$
is the action of the periodic orbit and $A$ is the sum of the areas of the two
triangles $OP_1'P_1$ and $OP_5P_7'$. \label{pomap}}
\end{figure}

An example of a trajectory $\alpha$ which has $k=5$ intersections
with the Poincar\'e surface in the encounter region, $P_1,\dots,P_5$, is shown
in figure \ref{pomap} (for positive $\Lambda_p$). Given such a trajectory one
can find a partner trajectory with $r$ more intersections in the encounter region
in the following way. Consider the line through the first point $P_1$ that is
formed by the intersections of trajectories that satisfy the required
initial conditions for scattering trajectories (indicated by the thin
line through $P_1$ in figure \ref{pomap}). There is second line through
the last point $P_k$ due to trajectories that satisfy the required final
conditions (the thin line through $P_5$ in the figure).
If one moves along the first line towards the stable manifold then one
finds for every $r>0$ a unique point $P_1'$ such that its $(k+r-1)$-th
iterate by the Poincar\'e map, $P_{k+r}'$, lies on the second line. 
This is the required partner
trajectory $\alpha'$. Figure \ref{pomap} shows an example for $r=2$. 
In the following we will show that one obtains the semiclassical 
periodic orbit contributions from this kind of trajectory pairs.
We first calculate the action difference of the trajectories. As a
consequence of the Poincar\'e-Cartan theorem the action integral
$\int \p \, \ud \q$ is independent of the path when it is evaluated
on a surface that is formed by a continuous family of trajectories
on the energy shell \cite{almeida88}. We apply this theorem to 
calculate three contributions to the action difference 
$\Delta S = S_{\alpha} - S_{\alpha'}$: $\Delta_i S$ coming
from the initial parts of the trajectories up to $P_1$ and $P_1'$,
$\Delta_f S$ coming from the final parts of the trajectories 
from $P_k$ and $P_{k+r}'$ onwards, and $\Delta_m S$ from the
remaining middle parts 
\begin{equation}
\Delta_i S = \int_{P_1'}^{P_1} p \ud q , \; \;
\Delta_f S = \int_{P_k}^{P_{k+r}'} p \ud q , \; \; 
\Delta_m S = \int_{P_1}^{O} p \ud q + 
             \int_{O}^{P_k} p \ud q - 
             \int_{P_1'}^{O} p \ud q -
             \int_{O}^{P_{k+r}'} p \ud q - r S_p .
\end{equation}
The first integral is evaluated along the line formed by trajectories
with the required initial conditions (they are all exponentially close to each
other at their starting points). The second integral is evaluated along the
line formed by trajectories with the required final conditions (they are all
exponentially close to each other at their end point). The remaining part of
the action difference is obtained by comparing the actions of the middle parts
of the trajectories to the action of the periodic orbit. The integrals are 
evaluated along lines that are formed by the intersection of trajectories
that continuously connect the middle parts of the trajectories to, respectively,
the $k$-fold and $(k+r)$-fold traversals of the primitive periodic orbit $p$.
If we add up all contributions we obtain an integral over the two triangular
regions $OP_1'P_1$ and $OP_kP_{k+r}$ depicted in figure~\ref{pomap}. The areas of
these regions are invariant under the transformation to normal form coordinates.
In the linearized approximation the sides of the triangles are straight lines.
If $P_1=(u,s)$ then $P_1' \approx (u \Lambda^{-r}_p,s)$,
$P_k \approx (u \Lambda^k_p,s \Lambda^{-k}_p)$,
$P_{k+r}' \approx (u \Lambda^k_p,s \Lambda^{-(k+r)}_p)$, and one obtains 
\begin{equation} \label{actdiff}
\Delta S = s u (1 - \Lambda_p^{-r}) - r S_p \; .
\end{equation}
We did the derivation for positive $\Lambda_p$, but the coordinates of the points
and (\ref{actdiff}) hold also for negative $\Lambda_p$. In the case that $\Lambda_p$
is negative and $r$ is odd then $P_1$ and $P_1'$ lie on different sides of the $s$-axis,
and $P_k$ and $P_{k+r}'$ on different sides of the $u$-axis.

Next we compare the semiclassical amplitudes of the trajectories $\alpha$
and $\alpha'$. They depend on the stability matrices of the trajectories.
The stability matrix describes the motion of neighbouring trajectories in
the linearized approximation. It gives the deviations at the final point
of a trajectory in terms of the deviations at the initial point, in coordinates
perpendicular to the trajectory
\begin{equation}
\begin{pmatrix}
\delta q_f \\ \delta p_f
\end{pmatrix}
= M
\begin{pmatrix}
\delta q_i \\ \delta p_i
\end{pmatrix} \; .
\end{equation}
For cavities with leads the amplitude of a scattering trajectory is proportional
to $|M_{21}|^{-1/2}$ where $M_{21}$ is a matrix element of $M$ \cite{rs02}.
The stability matrix can be composed by multiplying stability matrices
for different parts of the trajectory. For the $k$ iterations of the Poincar\'e map in the
vicinity of the periodic orbit the stability matrix of $\alpha$ is approximated
by the stability matrix of the periodic orbit $p$ and hence we write
$M_\alpha = M_f M_p^k M_i$ which $i$ and $f$ stand for the initial 
and final parts of the trajectory. The corresponding approximation for
the trajectory $\alpha'$ is $M_{\alpha'} = M_f M_p^{k+r} M_i$. 
Powers of the two-dimensional stability matrix $M_p$ can be written
in terms of their eigenvalues as
\begin{equation} \label{stretch}
M_p^{k} = \Lambda_p^k P_u + \Lambda_p^{-k} P_s
\sim \Lambda_p^k P_u
\quad \text{as} \quad k \to \infty \, ,
\end{equation}
where $P_s$ and $P_u$ are the projection operators onto the eigenvectors
of $M_p$ in the stable and unstable directions, respectively.
As a consequence $M_{\alpha'} \sim M_\alpha \Lambda^r_p$ as $k\to \infty$,
and we obtain the approximation 
\begin{equation} \label{amprel}
A_{\alpha'} \approx A_\alpha \, |\Lambda_p|^{-r/2} \; .
\end{equation}
It is convenient to have the following geometrical picture. One considers
a neighbouring trajectory with initial infinitesimal deviation
$\delta q_i \neq 0$ and $\delta p_i=0$ and follows the development of this
deviation in time as one moves along the trajectory. A conjugate point occurs
every time the projection of the deviation onto the (orthogonal) $p$-direction
is zero, and the index $\nu$ increases by one. (For billiard with Dirichlet
boundary conditions it increases also by two for every reflection on the wall.)
The matrix element $M_{21}$ is equal to the projection at the final point.
During the time the trajectory follows the periodic orbit, the deviation is
aligned and stretched along the unstable direction, according to (\ref{stretch}),
and rotates with it around the periodic orbit. The Maslov index of the
periodic orbit is the number of times the stable and unstable manifolds
rotate by half a turn \cite{crl90} (plus twice the number of reflections
on walls with Dirichlet b.c.s). After each traversal of the periodic orbit,
the manifolds are back where they started and the Maslov index is an integer.
As long as both $\alpha$ and $\alpha'$ are close to the periodic orbit, 
$\alpha'$ will pick up $r$ times the Maslov index of the orbit $p$
over  the trajectory $\alpha$. Outside of the encounter, both trajectories 
are close to each other and have the same index, giving
\begin{equation} \label{maslov}
\nu_{\alpha'} = \nu_{\alpha} + r \mu_{p} \; .
\end{equation}
We mention that this geometrical picture can be generalized to
systems with higher degrees of freedom, $f>2$. Then one considers the
development of an $(f-1)$-dimensional volume element in time that is
formed at the initial point by infinitesimal deviations in the $(f-1)$
$q$-directions orthogonal to the trajectory.

We can now put the trajectory pairs that follow a trapped periodic orbit 
$p$ in this way into the semiclassical sum for $C(\epsilon)$, 
equation~(\ref{cepssemi}). The times of the two trajectories are related
by $T_{\alpha'} \approx T_{\alpha} + r T_p$. We insert furthermore 
(\ref{actdiff}), (\ref{amprel}), and (\ref{maslov}) into (\ref{cepssemi})
and take into account a factor of two if the channels $a$ and $b$ are
the same in systems with time reversal symmetry, because then one can pair
a trajectory also with the time-reverse of its partner. This factor is
denoted as before by $\eta_{ab}$, see (\ref{etadef}). We are left
with the following to evaluate
\begin{equation} \label{sum_po1}
C^{p,r}(\epsilon) \approx \sum_{a,b} \eta_{ab}
\frac{\expp^{-\frac{\ci}{\hbar} r S_{p} 
+ \frac{\ci \pi}{2} r \mu_p}
\expp^{\frac{\ci \epsilon \mu}{2} r T_p}
}{T_{\mathrm{H}} \, |\Lambda_p|^{\frac{r}{2}}} 
\sum_{\alpha (a \to b)} |A_{\alpha}|^2
\, \expp^{\frac{\ci}{\hbar} s u (1 - \Lambda_p^{-r})} \,
\expp^{\ci \epsilon \mu T_\alpha} + (r \to -r) \; .
\end{equation}
The last term denotes the same contribution with $r$
replaced by $-r$. It is obtained from interchanging $\alpha$
and $\alpha'$ in the double sum over trajectories (\ref{cepssemi}).
The remaining step consists in performing the sum over all trajectories
$\alpha$ that enter the encounter region of the periodic orbit $p$. 
This summation is performed by applying the ergodicity property that very
long trajectories explore the available phase space volume uniformly,
combined with the finite escape probability of the scattering trajectories.
As a consequence the sum over trajectories, with the amplitudes
as weight factors, are replaced by an integral
(for related sum rules in closed systems see \cite{sie99})
\begin{equation} \label{replace}
\sum_{\alpha (a \to b)} |A_{\alpha}|^2 \ldots
\approx \int \ud T \, \int \ud s \, \ud u
\; w_{p,T}(s,u) \, \expp^{- \mu T_{\mathrm{exp}}} \ldots \; ,
\end{equation}
where $T$ is the total time of the trajectory ($=T_\alpha$) and $w_{p,T}$ is the
probability density that a trajectory of time $T$ goes through a surface element
$\ud s \, \ud u$ around the point $(u,s)$ in a fixed Poincar\'e section transverse
to the periodic orbit $p$. The integrals over $s$ and $u$ are limited by the constant
$c$. The time $T_{\mathrm{exp}}$ is the time during which the orbit can escape. It differs
from the time $T$ of the trajectory, because the trajectory cannot escape during
the time it follows the trapped periodic orbit in the encounter region of the orbit
(if one chooses the arbitrary constant $c$ sufficiently small). Hence, the
difference between $T$ and $T_{\mathrm{exp}}$ is the time in the encounter
region which is given by
\begin{equation} \label{enctime2}
t_{\mathrm{enc}}^p \approx k \, T_p \approx \frac{1}{\lambda_p} 
\ln \frac{c^2}{|u s|} \, ,
\end{equation}
where $k$ is the number of iterations in the encounter region, and
the Lyapunov exponent $\lambda_p$ of the periodic orbit is obtained from 
$|\Lambda_p| = \expp^{\lambda_p T_p}$. The number $k$ was obtained
by using (\ref{map}) from which follows that 
$k \approx \frac{1}{\lambda_p T_p} \ln \frac{c}{|u|} +
\frac{1}{\lambda_p T_p} \ln \frac{c}{|s|}$.
It is remarkable that (\ref{enctime2}) has the same form as
the encounter time for self-encounters (\ref{enctime1}). For this 
reason, the central relation
(\ref{suinteqn}) for the integrals over $s$ and $u$ holds as well
and will be used in the following.

The probability density $w_{p,T}(s,u)$ is obtained by noting that
the probability that an element of a trajectory of time $\ud t$ enters the
surface element $\ud s \, \ud u$ is given by
\begin{equation}
\frac{\ud s \, \ud u \, \ud t}{\Omega} \; .
\end{equation}
The probability density follows as
\begin{equation} \label{probdens}
w_{p,T}(s,u)=\int_{0}^{T-t_{\mathrm{enc}}^p}
\ud t_{1} \frac{T_{p}}{\Omega \, t_{\mathrm{enc}}^p} \, ,
\end{equation}
where $t_1$ is the time at the intersection with the Poincar\'e surface.
Equation (\ref{probdens}) contains the factor
$T_p/t_{\mathrm{enc}}^p \approx 1/k$ to remove the multiple counting
of a trajectory, because $k$ points in the Poincar\'e section correspond
to the same trajectory.
We substitute (\ref{replace}) and (\ref{probdens}) into (\ref{sum_po1})
and change the integration variable from $T$ to $t_2$, where $t_2$
is the time from the encounter region to the exit channel,
$T=t_1+t_{\mathrm{enc}}^p+t_2$, and obtain
\begin{equation}
C^{p,r}(\epsilon) \approx \sum_{a,b} \eta_{ab}
\frac{T_p \expp^{-\frac{\ci}{\hbar}r S_{p} + \frac{\ci \pi}{2} r \mu_{p}}
\expp^{\frac{\ci \epsilon \mu}{2} r T_p}
}{T_{\mathrm{H}} \, |\Lambda_p|^{\frac{r}{2}}} \;
\int \ud  s \, \ud u \frac{\expp^{\frac{\ci}{\hbar} s u (1 - \Lambda_p^{-r})}
\expp^{\ci \epsilon \mu t_{\mathrm{enc}}^p}
}{\Omega \, t_{\mathrm{enc}}^p}
\prod_{i=1}^2 \int_{0}^{\infty} \ud t_{i} \;
\expp^{-\mu (1 - \ci \epsilon) t_i} + 
(r \to -r).
\end{equation}
We see again that the expression factorizes into contributions from the
links and the encounter region. This implies that the diagrammatic
rules can be extended, and that one gets additional contributions
from the periodic orbit encounters. We can now perform the integral
over $s$ and $u$. Note that this integral automatically sums over
trajectories $\alpha$ with an arbitrary number $k$ of windings around
the periodic orbit. We expand the integrand in $\epsilon$ up to first
order and use equation (\ref{suinteqn}). Only the term with
$t_{\mathrm{enc}}^p$ in the numerator cancels with the encounter time
in the denominator and contributes semiclassically. After integrating
there is again a cancellation of Heisenberg times and the diagrammatic
rule for the $r$-th repetition of a periodic orbit $p$ follows as
\begin{equation} \label{diapo}
2 \ci \epsilon \mu A_{p,r} 
\cos\left(-\frac{1}{\hbar}r S_{p} + \frac{\pi}{2} r \mu_{p}
+ \frac{\epsilon \mu}{2} r T_p \right)
\end{equation}
where we used the identity  $\sqrt{|\det(M_p^r - 1)|}=|\Lambda_p|^{r/2}
|1 - \Lambda_p^{-r}|$. The semiclassial contribution (\ref{diapo}) comes
from all the trajectory pairs for which one trajectory winds $r$ more
times around the periodic orbit $p$ than its partner. The contribution
comes from the close vicinity of the origin, i.e.\ from trajectory
pairs for which the {\em total} number of windings around the
periodic orbit is very large. The diagrammatic rules that we have
encountered so far are the first three in table~\ref{table}, and the 
total contribution to $C(\epsilon)$ follows as
\begin{equation}
C^{p,r}(\epsilon) \approx \sum_{a,b} \eta_{ab} 
\frac{2 \ci \epsilon \mu}{M^2 \, (1 - \ci \epsilon)^2} A_{p,r} 
\cos\left( -\frac{1}{\hbar}r S_{p} + \frac{\pi}{2} r \mu_{p} 
+ \frac{\epsilon \mu}{2} r T_p
\right) \; .
\end{equation}
Finally we perform the sum over the channels which gives a factor
$M (M + \kappa - 1)$ as in section~\ref{average}. The result is
\begin{equation}
C^{p,r}(\epsilon) \approx 2 i \epsilon \mu \frac{M(M + \kappa - 1)}{M^2}
A_{p,r} \cos\left( -\frac{1}{\hbar}r S_{p} + \frac{\pi}{2} r \mu_{p} \right)
+ O(\epsilon^2) \; .
\end{equation}
In the case of systems without time-reversal symmetry ($\kappa=1$) this yields
exactly the contribution of the $r$-th repetition of the trapped periodic
orbit to the time delay. However, for systems with time-reversal symmetry
the prefactor is slightly wrong. The reason for this is the same as why
the diagonal approximation in section~\ref{average} did not give the 
correct mean time delay. There are other correlations between trajectories
that have to be included. So far we have considered trajectory pairs, $\alpha$
and $\alpha'$, which differ only in the number of windings around the 
periodic orbit $p$. However, the orbit $\alpha$ can also have additional
close self-encounters. As a consequence, $\alpha$ and $\alpha'$ can differ
also in the way in which different links are connected in the encounter regions
of the self-encounters. We will see in the next section that this leads
to a variety of further correlations.

\subsection{Combinations of periodic orbit encounters and self-encounters}
\label{subsect2}

Consider a trajectory pair with self-encounters. The number and types of
the self-encounters are specified by the vector $\v$. In addition, the
trajectory pair can have an encounter with a periodic orbit. We consider first
the situation where the periodic orbit encounter occurs during one of the
$(L+1)$ links. The additional encounter divides a link into two parts and
increases the total number of links to $(L+2)$. The joint probability density
for this case is
\begin{equation} \label{wvpt}
w_{\v,p,T}(\s,\u)=\int' \ud t_1 \ldots \ud t_{L+1} \; \frac{T_p}{\Omega^{L+1-V}
t_{\mathrm{enc}}^p \prod_{\sigma=1}^{V} t_{\mathrm{enc}}^{\sigma}} \; .
\end{equation}
where the prime denotes again that all link times, including the
remaining one $t_{L+2}$, have to be positive (see discussion around
equation (\ref{tt})).
With this probability density one finds again that the contribution
to the correlation function $C(\epsilon)$ factorizes and that it can
be obtained by the diagrammatic rules that have been derived in the
previous sections. They are summarized in the first three lines of
table~\ref{table}. After a summation over the channels one obtains
\begin{equation} \label{joint1}
C^{\v,p,r}_{\mathrm{I}}(\epsilon) \approx (L+1) \, \frac{M (M + \kappa - 1)}{M^{L-V+2}}
N(\v) (-1)^V \, 2 i \epsilon \mu  A_{p,r}
\cos\left( -\frac{1}{\hbar}r S_{p} + 
\frac{\pi}{2} r \mu_{p} \right) + O(\epsilon^2) \; .
\end{equation}
There is a factor $(L+1)$, because the periodic orbit encounter can occur
during any of the original $(L+1)$ links.

\begin{figure}
\begin{center}
\mbox{\epsfxsize6cm\epsfbox{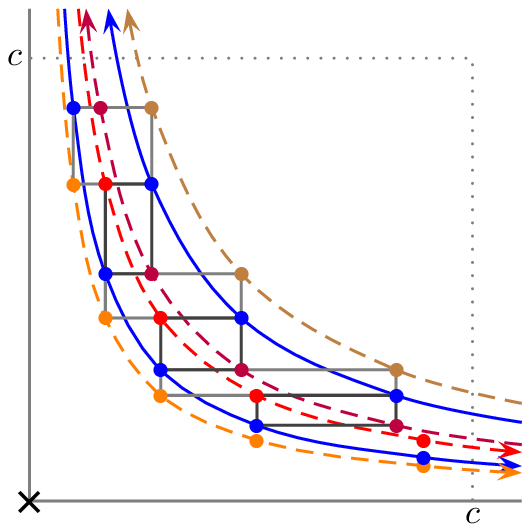}}
\end{center}
\caption{An example of a trajectory (the dots on the two full lines)
with a two-encounter in the vicinity of a periodic orbit and some of
its partner trajectories (the dots on the dashed lines).
The partner trajectories are obtained by drawing rectangular boxes which
have points of the original trajectory (full lines) in opposite corners. \label{orbenc4}}
\end{figure}

One can also have the situation that a periodic orbit encounter overlaps
with a self-encounter. In other words a self-encounter occurs in the vicinity
of a periodic orbit. This leads to interesting consequences.
In the simplest situation a two-encounter occurs in the vicinity of a
periodic orbit in a system with time-reversal symmetry. A Poincar\'e
section for this case is shown in figure~\ref{orbenc4}. An orbit $\alpha$,
indicated by the full lines, has a first encounter with the periodic orbit,
and its intersection points with the Poincar\'e section follow a hyperbola
in the downward direction (the direction is indicated by arrows).
Then the trajectory goes away from the periodic orbit, makes a loop and
comes back and follows the periodic orbit in the opposite direction. The
(time-reverses of) the intersection points with the Poincar\'e surface
then follow a second hyperbola in the upward direction. (When we speak
of the intersection points of $\alpha$ in the following, we mean this
to include the intersection points of its time reverse.)

This situation is different from a usual two-encounter where $\alpha$
has only two intersection points with the Poincar\'e surface, $(u_1,s_1)$
and $(u_2,s_2)$. In a usual two-encounter the partner orbit $\alpha'$
has intersection points which are approximately given by $(u_1,s_2)$
and $(u_2,s_1)$, i.e. they are obtained by drawing a rectangle with the
intersection points of $\alpha$ in opposite corners.
If we apply this picture to find the partner orbit in figure~\ref{orbenc4}
we see that we have several possibilities to use this construction to find a
partner orbit which traverses the loop in the opposite direction, because
$\alpha$ has many intersection points. Some of these possibilities lead to the
same partner trajectory. For example, in figure~\ref{orbenc4} the rectangles
that are between the two full lines yield a partner trajectory that follows
the two dashed hyperbolae that lie between the two full hyperbolae. 
If other intersection points of $\alpha$ are connected by rectangles, then
other partner trajectories are obtained. Figure~\ref{orbenc4} shows one
further example by the dashed hyperbolae that lie outside the full hyperbolae.
But there are more possibilities that are not shown in the figure. The
number $k$ of rectangles that correspond to the same partner trajectory
follow from the Poincar\'e mapping (\ref{map}) with $|\Lambda_p|=
\expp^{\lambda_p T_p}$. It has the form
$k = \frac{1}{\lambda_p T_p} \ln \frac{c}{\max_j |u_j|} +
    \frac{1}{\lambda_p T_p} \ln \frac{c}{\max_j |s_j|}$.
This suggests the following definition of the encounter time
\begin{equation} \label{enctime3}
t_{\mathrm{enc}}^{p,\sigma} \approx k T_p \approx \frac{1}{\lambda_p} \ln \frac{c^2}{\max_i |s_i|
\times \max_j |u_j|} \, ,
\end{equation}
which has a close similarity to (\ref{enctime1}).

The occurrence of several partner orbits can be understood in the following
way. A partner trajectory $\alpha'$ of $\alpha$ differs in the direction in
which the loop between the two periodic orbit encounters is traversed. But it
can also differ in the number of times it follows the periodic orbit
before and after the loop, as long as the total number of traversals is the
same as for $\alpha$. For example if $\alpha$ has $k_1$ intersections in
the square of the Poincar\'e section with side lengths $c$ before the loop,
and $k_2$ intersections after the loop, then the partner trajectory can have
$k_1+d$ intersections before the loop and $k_2-d$ intersections after the loop,
where $d$ is any integer with $k_1+d >0$ and $k_2-d>0$. This means that
the presence of the periodic orbit leads to a large number of possible
partner trajectories. 

However, we are interested in obtaining the periodic orbit contributions
to the time delay. Hence we have to consider trajectories $\alpha''$ which
have altogether $r$ more periodic traversals than the trajectories $\alpha'$.
These trajectories $\alpha''$ then differ from the original trajectories $\alpha$
in the number of periodic orbit traversals as well as in the direction
in which the loop is traversed. In the following we calculate the
semiclassical contribution of the trajectory pairs $\alpha$ and $\alpha''$.

Let us consider the general case that an $l$-encounter occurs near
a periodic orbit $p$. A partner orbit $\alpha'$ is obtained by reconnecting the
links in a different way in the encounter region. This reconnection
is specified by a permutation $\pi(j)$, $j=1,\ldots,l$ which consists
of one cycle of length $l$ \cite{muller04,muller05b}. The action
difference in \cite{muller04,muller05b} can be expressed in the form
$\Delta S = \sum_{j=1}^l s_j (u_j - u_{\pi(j)})$. If we now consider
partner trajectories $\alpha''$ that have additional periodic orbit traversals
$r_1,\ldots,r_l$ during the $l$ encounters with the periodic orbit, with
$\sum_j r_j = r$, then the action difference with $\alpha$ changes according
to (\ref{actdiff}) and is given by
\begin{equation} \label{actdiff2}
\Delta S =  \sum_{j=1}^l s_j (u_j - u_{\pi(j)} \Lambda_p^{-r_j}) - r S_p = \s^{\mathrm{T}} B \u - r S_p
\end{equation}
where $\s$ and $\u$ are the vectors of the $l$ coordinates in the stable and
unstable directions, and $B= I + \tilde{B}$ where $\tilde{B}_{ji} = - \delta_{i \pi(j)}
\Lambda_p^{-r_j}$. We have $\det B = 1 + \det \tilde{B}$, because the permutation
consists of one cycle of length $l$. Furthermore, $(l-1)$ column exchanges bring $\tilde{B}$
into diagonal form, and hence $\det \tilde{B} = - \Lambda_p^r$, and
$\det B = 1 - \Lambda_p^r$.

If we have one $l$-encounter $\sigma$ in the vicinity of a periodic orbit $p$, and no
further self-encounters, then the probability density is given by
\begin{equation} \label{wspt}
w_{\sigma,p,T}(\s,\u)=\int' \ud t_1 \ldots \ud t_l \; \frac{T_p}{\Omega^{l}
t_{\mathrm{enc}}^{p,\sigma}} \; .
\end{equation}
where the encounter time is given in (\ref{enctime3}). The factor $1/k=T_p/t_{\mathrm{enc}}^{p,\sigma}$
takes care of the overcounting of trajectory pairs. The contribution to $C(\epsilon)$ follows
from (\ref{sum_po1}), (\ref{replace}) and (\ref{wspt}) as
\begin{equation} \label{calc}
\sum_{a,b} \eta_{ab}
\frac{T_p \expp^{-\frac{\ci}{\hbar} r S_{p} + \frac{\ci \pi}{2} r \mu_{p}}
\expp^{\frac{\ci \epsilon \mu}{2} r T_p}
}{T_{\mathrm{H}} \, |\Lambda_p|^{\frac{r}{2}}} \;
\int \ud  \s \, \ud \u \frac{\expp^{\frac{\ci}{\hbar} \s^{\mathrm{T}} B \u}
\expp^{\ci l \epsilon \mu t_{\mathrm{enc}}^{p,\sigma}}
}{\Omega^l \, t_{\mathrm{enc}}^{p,\sigma}}
\prod_{i=1}^{l+1} \int_{0}^{\infty} \ud t_{i} \;
\expp^{-\mu (1 - \ci \epsilon) t_i} + 
(r \to -r).
\end{equation}
The expression factorizes again, and the contribution
from the encounter region is obtained by expanding the exponential which
contains the encounter time. The contribution originates again from the
linear term of this expansion. The resulting diagrammatic rule is listed
in the forth line of table \ref{table}.
Note that the contribution does not depend on how the
repetition number $r$ is split into parts $r_1,\ldots,r_l$. In fact, the integral
over the $\s$ and $\u$ coordinates sums over the different ways of splitting $r$
into parts, because a trajectory $\alpha$ has many intersection points. The different
$\s$ and $\u$ coordinates for these intersection points correspond to different
ways of splitting $r$, as in figure \ref{orbenc4} and the discussion after
equation (\ref{enctime3}).
\begin{table}
\begin{center}
\begin{tabular}{lc} \hline \\[-2mm]

contribution of each link &
$\frac{1}{M (1 - i \epsilon)}$ \\[2mm]

contribution of each $l$-encounter & 
$-{M (1 - i l \epsilon)}$ \\[2mm]

$r$-th contribution of a periodic-orbit encounter &
$2 \ci \epsilon \mu
A_{p,r} \cos\left( -\frac{1}{\hbar}r S_{p} + 
\frac{\pi}{2} r \mu_{p} + \frac{1}{2} \epsilon \mu r T_p \right)$ \\[2mm]

$r$-th contribution of periodic-orbit plus $l$-encounter \; \; \; &
$2 \ci l \epsilon \mu
A_{p,r} \cos\left( -\frac{1}{\hbar}r S_{p} + 
\frac{\pi}{2} r \mu_{p} + \frac{1}{2} \epsilon \mu r T_p \right)$ \\[1mm]
\hline
\end{tabular}
\end{center}
\caption{Diagrammatic rules for the different contributions
to the correlation function $C(\epsilon)$.}
\label{table}
\end{table}

Finally, we include the possibility that there are additional self-encounters
which are not near a periodic orbit. One finds that also in this case the
contributions factorize, and they can be evaluated by the rules in table~\ref{table}.
If the total number of self-encounters is given by the vector $\v$, and we 
sum over all the cases where one of the $V$ self-encounters is in the
vicinity of the periodic orbit $p$, we obtain
\begin{equation} \label{joint2}
C^{\v,p,r}_{\mathrm{II}}(\epsilon) \approx 
\left(-\sum_\sigma l_\sigma \right)
\frac{M (M + \kappa - 1)}{M^{L-V+2}} N(\v) (-1)^V \,
2 i \epsilon \mu  A_{p,r} \cos\left( -\frac{1}{\hbar}r S_{p} + 
\frac{\pi}{2} r \mu_{p} \right) + O(\epsilon^2) \; .
\end{equation}
This expression differs from (\ref{joint1}) in that the factor $(L+1)$ is replaced
by $-\sum_\sigma l_\sigma=-L$. When we add the two this factor becomes one.
Finally we sum over all vectors $\v$ and include also the contributions from
section \ref{subsect1} (corresponding to $L=V=0$ and $N(\v)=1$).
\begin{align}
C^{p,r}(\epsilon) & \approx (M + \kappa - 1) \sum_{k=1}^{\infty}
\sum_{\v}^{L-V+1=k} \frac{N(\v) (-1)^V}{M^k}
2 i \epsilon \mu  A_{p,r} \cos\left( -\frac{1}{\hbar}r S_{p} + 
\frac{\pi}{2} r \mu_{p} \right) + O(\epsilon^2)
\notag \\ \label{final} &
\approx  2 i \epsilon \mu  A_{p,r} \cos\left( -\frac{1}{\hbar}r S_{p} + 
\frac{\pi}{2} r \mu_{p} \right) + O(\epsilon^2) \; .
\end{align}
where the sum rule (\ref{recursion}) for $N(\v)$ has been used.

Equation (\ref{final}) is the final result of this section.
Using (\ref{timedelaytrajsym}), it gives the correct contribution of
the $r$-th repetition of the periodic orbit $p$ to the time delay
(\ref{timedelayeq}) in systems with or without time reversal symmetry.

We have seen in this section that the vicinity of a periodic orbit
gives rise to a rich variety of correlations between trajectories.
We concentrated on those correlations that are responsible for the
semiclassical periodic orbit terms of the time delay. There are
further correlations which we do not explore in the present article.
One example are the trajectory pairs of this section with $r=0$.
There can also be multiple encounters with a periodic orbit $p$
which do not correspond to the picture that one of the usual
self-encounters occurs near a periodic orbit. An example is a
trajectory which visits a periodic orbit twice and follows it
both times in the {\em same} direction and has no further self-encounters.
Furthermore, a trajectory can have encounters with several different
periodic orbits. These terms, however, appear in higher order terms
in the expansion in $\epsilon$. (Visits of $n$ periodic orbits appear
in terms of order $\epsilon^n$.) These additional correlations merit
further study.

An interesting question is whether one obtains periodic orbit contributions
to the conductance with the type of open orbit correlations in this section.
The Landauer-B\"uttiker formulation for the conductance can be obtained,
in certain situations, from linear response theory (Kubo formula). On the
other hand, when the Kubo approach is applied to the bulk conductivity in
antidot lattices it gives semiclassical expressions in terms of periodic
orbits \cite{richter00}. It is an open question whether periodic orbit
contributions exist for the Landauer-B\"uttiker conductance as well.

The conductance is proportional to the total transmission through the cavity
\begin{equation}
{\cal T} = \sum_{b_{\mathrm{out}} a_{\mathrm{in}}} S_{b_{\mathrm{out}} a_{\mathrm{in}}}(E)
S_{b_{\mathrm{out}} a_{\mathrm{in}}}^*(E) \; ,
\end{equation}
where the only difference to $C(\epsilon)$ lies in the different channel sum (the sum
is over incoming and outgoing channels), and $\epsilon = 0$. Since the periodic orbit
contributions to $C(\epsilon)$ in (\ref{final}) vanish in the limit $\epsilon = 0$
one finds that there is no periodic orbit contribution to the conductance as well
in this leading order semiclassical calculation. This does not completely rule out
the possibility of periodic orbit terms in the conductance, but if they exist they
need to have a different sign in the reflectance, because transmission and reflectance
have to add up to a constant, the number of channels in the incoming lead(s).

\section{Correlation functions of the time delay} \label{correl}

In this section we look at a correlation function of the time delay, and show that it
is possible to get the leading order result of RMT using open trajectories. This addresses
the concern of \cite{lv04} that the diagonal approximation in the approach using trajectories 
gave a different result.  The key is that there are other contributions of the same order 
as the diagonal approximation for open trajectories, and these need to be included to 
get the leading order term.  The calculation for the correlation function of the
time delay is similar to that for the conductance variance and, in particular,
to the Ericson fluctuations \cite{muller07}. For this reason we shall be brief
and only sketch the calculations. The correlation function that we consider
is defined as
\begin{equation} \label{taucorr}
\tilde{R}_{2}(\omega,M) = \frac{\left\langle
\tau_{\mathrm{W}} \left(E + \frac{\omega M}{4 \pi \bar{d}}\right) 
\tau_{\mathrm{W}} \left(E - \frac{\omega M}{4 \pi \bar{d}}\right)
\right\rangle_E - \bar{\tau}_{\mathrm{W}}^2}{\bar{\tau}_{\mathrm{W}}^2} \, .
\end{equation}
Again it is convenient to specify the energy difference in units of
$M (2 \pi \bar{d})^{-1}$. The correlation function involves an energy
average over an energy range $\Delta E$ which is classically small,
but encompasses many resonances $E>>\Delta E >>1/\bar{d}(E)$.  
In RMT the leading order result for a large number of channels $M$
is given for the two considered symmetry classes by \cite{eckhardt93, val98}
\begin{equation} \label{corrfnexp}
\tilde{R}_{2}(\omega,M) =\frac{2 \kappa}{M^2}
\frac{1 - \omega^2}{(1 + \omega^2)^2} \, .
\end{equation}
When we express the correlation function in (\ref{taucorr}) as a sum over
quadruples of trajectories we subtract the term $\bar{\tau}_{\mathrm{W}}^2$
by removing the trajectories pairs of section~\ref{average} that give the
mean delay time. Inserting the semiclassical approximation for the time delay
(\ref{timedelaytraj}) results in
\begin{equation} \label{corrfntraj}
\tilde{R}_{2}(\omega,M) = \frac{1}{T_{\mathrm{H}}^{4}}\left\langle \sum_{\substack{a,b \\ c,d}}
\sum_{\substack{\alpha,\alpha' (a\to b) \\ \beta,\beta' (c\to d)}}^{\prime}
T_{\alpha} T_{\beta} A_{\alpha}A^{*}_{\alpha'}A_{\beta}A^{*}_{\beta'}
\expp^{\frac{\ci}{\hbar}(S_{\alpha}-S_{\alpha'}+S_{\beta}-S_{\beta'})}
\expp^{\frac{\ci \omega \mu}{2}(T_{\alpha}-T_{\alpha'}-T_{\beta}+T_{\beta'})} \right\rangle_{E} \, .
\end{equation}
Here, $\alpha,\alpha'$ are trajectories from channel $a$ to $b$ and $\beta,\beta'$ 
are trajectories from channel $c$ to $d$ and we sum over trajectories and channels.  
The prime at the second sum indicates that we have removed the orbits where
$\alpha\approx\alpha'$ and $\beta\approx\beta'$. 

We start with the calculation of the diagonal contribution from  \cite{lv04}.
Because we have removed the trajectory pairs that give the average time delay,
we have to consider only the cases when $S_{\alpha}=S_{\beta'}$ and
$S_{\beta}=S_{\alpha'}$. Without time reversal symmetry this means that
$\alpha=\beta'$ and $\beta=\alpha'$, which requires that the channels $a=c$
and $b=d$. Hence we get a factor $M^{2}$ from the sum over the channels.
With time reversal symmetry there are three additional cases. We can
also have $\alpha=\overline{\beta'}$ and $\beta=\overline{\alpha'}$, where
the overline indicates the time reverse. This requires $a=d$ and $b=c$ and
the channel sum gives  $M^2$. Furthermore, we can have $\alpha=\beta'$ and
$\beta=\overline{\alpha'}$, or $\alpha=\overline{\beta'}$ and $\beta=\alpha'$.
Both cases require $a=b=c=d$ and each channel sum yields $M$. The total channel
factor for the systems with time reversal symmetry is thus $2M(M+1)$, and in
leading order for large $M$ the channel factor for the two symmetry cases is
$\sim \kappa M^{2}$. We can perform now the sums over the trajectory pairs by
replacing them by an integral according to (\ref{opensumrule}), and we perform
the sum over the channels by multiplying with the channel factor
\begin{equation}
\tilde{R}_{2}^{\mathrm{diag}}(\omega,M) \approx 
\frac{\kappa M^{2}}{T_{\mathrm{H}}^{4}}\int_{0}^{\infty}
\int_{0}^{\infty}\ud T_{\alpha}\ud T_{\beta}\:T_{\alpha}T_{\beta}
\expp^{-\mu(T_{\alpha}+T_{\beta})}\expp^{\ci \omega \mu (T_{\alpha}-T_{\beta})} \, ,
\end{equation}
which, after integrating, is
\begin{equation}
\tilde{R}_{2}^{\mathrm{diag}}(\omega,M) \approx \frac{\kappa}{M^2}
\frac{1}{(1 + \omega^2)^2} \; .
\end{equation}
We have a factor of two different from the expected result (\ref{corrfnexp}), 
and a different functional form. To correct for this we have to add sums over
quadruplets with encounters that contribute at the same order of $M^{-1}$. The
reason why they can contribute at the same order as the diagonal term is because
they have larger channel factors.

\begin{figure}
\begin{center}
\mbox{\epsfxsize6cm\epsfbox{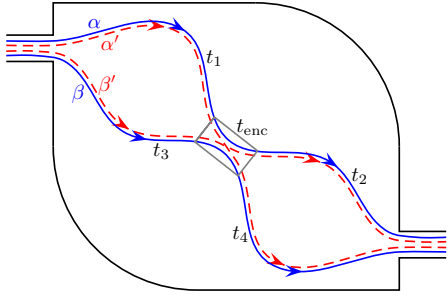}}
\end{center}
\caption{An example of a trajectory quadruplet where the trajectories $\alpha$ and
$\beta$ (full lines) have one two-encounter. Their partner trajectories (dashed
lines) cross the encounter region differently. The encounter region is
indicated by a rectangular box. \label{trajquad}}
\end{figure}
One example is a quadruplet where the trajectories $\alpha$ and $\beta$ have one
close encounter. Then their partner trajectories follow one of these two trajectories
from the entrance channel to the encounter region where they switch over to the
other trajectory and follow it to the exit channel, see figure \ref{trajquad}
(or figure 4j in \cite{muller07}).
If we label the link time betweens entrance or exit channels and the encounter 
region by $t_i$, then the times along $\alpha$ and $\beta$ can be written as
\begin{equation} 
T_{\alpha} = t_{1} + t_{\mathrm{enc}} + t_{2} \, , \quad 
T_{\beta}  = t_{3} + t_{\mathrm{enc}} + t_{4} \, .
\end{equation}
If $\alpha'$ follows first $\alpha$ and then $\beta$, and $\beta'$ does vice versa,
we have
\begin{equation} \label{case1}
T_{\alpha'} = t_{1} + t_{\mathrm{enc}} + t_{4} \, , \quad
T_{\beta'}  = t_{3} + t_{\mathrm{enc}} + t_{2} \, .
\end{equation}
These quadruplets are only possible if channels $b$ and $d$ are identical. We can
also interchange the role of $\alpha'$ and $\beta'$ which requires that $a=c$.
In systems with time reversal symmetry we have to consider also cases where the
role of $\alpha'$ and/or $\beta'$ is taken over by its time reverse, similarly as
for the diagonal approximation. The sum over trajectories for these different
configurations always gives the same result, hence it is sufficient to do the
calculation for the case (\ref{case1}) and to perform the channel sum by multiplying
with the total channel factor. In systems without time reversal symmetry this
channel factor is $2 M^3$, and in system with time reversal symmetry it is 
$4 M^3 + 8 M^2 + 4 M$. In leading order it is hence $\sim 2 \kappa M^3$.

The difference in times for the case (\ref{case1}) is
\begin{equation}
\frac{1}{2}(T_{\alpha}-T_{\alpha'}-T_{\beta}-T_{\beta'})=t_{2}-t_{4} \, ,
\end{equation}
and the contribution to the correlation function is given by
\begin{align}
\nonumber 
&\frac{2 \kappa M^{3}}{T^{4}_{\mathrm{H}}}
\int_{0}^{\infty}\ud t_{1}\ud t_{2}\ud t_{3} \ud t_{4}\:
\expp^{-\mu(t_{1}+t_{2}+t_{3}+t_{4})}
\expp^{\ci \omega \mu (t_{2}-t_{4})} \\
& \qquad \times \int \ud s \ud u \:
(t_{1}+t_{\mathrm{enc}}+t_{2}) (t_{3}+t_{\mathrm{enc}}+t_{4})
\frac{\expp^{-\mu t_{\mathrm{enc}}}}{\Omega t_{\mathrm{enc}}}
\expp^{\frac{\ci}{\hbar} s u} \, .
\end{align}
This then can be evaluated with the rule (\ref{suinteqn}) and yields
\begin{equation}
\frac{2 \kappa}{M^2} \frac{\omega^2}{(1 + \omega^2)^2} \, .
\end{equation}
Following a similar process we find the contribution from the other diagrams in 
figure~4 of \cite{muller07} and give the results in table~\ref{trajtabletimecorrnonsym}.
\begin{table}[htb]
\centering
\begin{tabular}{|c|c|c|}
\hline
4c&4d&4e\\
$\frac{-1 - 2 \omega^2}{M^6 (1 + \omega^2)^2}$ &
$\frac{8 \omega^2 (1 + \omega^2)}{M^6 (1 + \omega^2)^2}$ &
$\frac{2 - 2 \omega^2 - 8 \omega^2 (1 + \omega^2)}{M^6 (1 + \omega^2)^2}$ \\
\hline
\end{tabular}
\caption{Contribution of different types of trajectory quadruplets to the time 
delay correlation function. The labelling corresponds to figure~4 in \cite{muller07}.}
\label{trajtabletimecorrnonsym}
\end{table}
When we now multiply these contributions by their channel factor $M^{4}+M^2 \sim M^4$ 
for the unitary case and add the terms calculated above, we get the following result of
\begin{equation}
\frac{2}{M^2}\frac{1 - \omega^2}{(1 + \omega^2)^2} \, ,
\end{equation}
which is indeed the leading order term of equation~(\ref{corrfnexp}).  For the 
orthogonal case, the channel factor for the configurations in table~\ref{trajtabletimecorrnonsym}
is $M^4+2M^3+3M^2+2M \sim M^{4}$ to leading order.  However there are additional
trajectory quadruplets that contribute (4f-4h) which are related to the quadruplets
4c-4e by time reversal of parts of the structure.  In fact these additional
quadruplets just give the same contribution again as those in figures 4c-4e.
Therefore, the leading order contribution is simply twice that for the unitary
case, again in line with equation~(\ref{corrfnexp}).

It is worth noting here that using the symmetrized version of the time delay,
equation~(\ref{timedelaytrajsym}) with (\ref{cepssemi}), we get a different result for
each trajectory quadruplet, but the sum of their contributions gives the same
result as here. To calculate higher order terms, using the methods of
\cite{muller07}, the similarities can be exploited by defining a `symmetrized'
correlation function which can be written in terms of the correlation function
of the scattering matrix elements.
\begin{equation} \label{correlsymm}
\tilde{R}_{2}(\omega,M) = \frac{-1}{M^2}\frac{\ud^{2}}{\ud \epsilon_1 \ud \epsilon_2}
\left\langle C^{\mathrm{fl}}\left(\epsilon_1, E + \frac{\omega M}{4 \pi \bar{d}} \right)
             C^{\mathrm{fl}}\left(\epsilon_2, E - \frac{\omega M}{4 \pi \bar{d}} \right) 
\right\rangle_E \Big\vert_{\epsilon_1=\epsilon_2=0}
\end{equation}
where $C^{\mathrm{fl}}(\epsilon, E)$ is the semiclassical approximation to the 
fluctuating part of the correlation function of the scattering matrix elements at energy $E$.
This means that the trajectory pairs responsible for the average part are removed.
The calculation is very similar to that of the Ericson fluctuations.
A complication in comparison to the calculation of the mean time delay
in chapter~\ref{average} is that one does not have one simple diagrammatic
rule for every link. Each link is traversed by two of the four trajectories
$\alpha$, $\alpha'$, $\beta$, $\beta'$, and the diagrammatic rule depends
on which two of the four trajectories that are. The summation over the
contributions from the different {\em families} is best done by a computer.

A final point is that if we consider 
$C^{\mathrm{fl}}(0,E)\, C^{\mathrm{fl}}(0, E)$ we are calculating
\begin{equation}
\left(\Tr\left[S(E) S^{\dagger}(E)\right]-M\right)^{2}
\end{equation}
which should be zero because of the unitarity of the scattering matrix. 
We checked this by using the formulae for the conductance variance
\cite{muller07}, but with the appropriate channel factors that are
given in this section.  The result is indeed $0$, and the unitarity
of the scattering matrix is preserved by the semiclassical
approximation if all semiclassical contributions are included.

\section{Conclusions} \label{concl}

The time delay is an interesting quantity to study because of the two
alternative semiclassical descriptions of it. The picture in terms of
scattering trajectories is similar to the conductance, and by using the
semiclassical methods of \cite{muller07} which includes trajectories with
self-encounters we obtained the average time delay. We considered 
also a correlation function of the time delay. Here the diagonal
approximation is not enough to obtain the leading order term (as noted
by \cite{lv04}), because more complicated trajectory quadruplets with
encounters contribute at the same order. By including these contributions
we find that the unitarity of the semiclassical scattering matrix is restored,
and that the semiclassical method does indeed provide an accurate description
in agreement with RMT. This is as expected from other correlation functions in \cite{muller07}.

The main result of this article, however, is the discovery of a new type of
scattering trajectory correlations that recreate the periodic orbit terms of
the time delay. These trajectories approach a periodic orbit, follow it closely
some number of times, and then part from it. Approaching the orbit guarantees
that they remain in the system, as long as the periodic orbit does and they remain
close to it. This is the reason why only the trapped periodic orbits appear in the
orbit sum of the time delay. For systems without time-reversal symmetry these
periodic orbit encounters are enough to recreate the periodic orbit terms
of the time delay, but for systems with time-reversal symmetry a small
additional contribution is needed. We obtained this contribution from combinations
of self-encounters and periodic orbit encounters.

We found that the occurrence of self-encounters in the close vicinity of a
periodic orbit leads to a wealth of new possible correlations between trajectories.
We considered in this article only those correlations that are needed for the
periodic orbit terms of the time delay, but there are further possible correlations.
These new type of correlations deserve further study, and it can be expected that they
play a role also in other contexts. For example, they might be relevant for
periodic orbit correlations in closed systems, as envisioned in \cite{almeida89}.
For the particular case of the Landauer-B\"uttiker conductance, however, we found
that the trajectory correlations
that yield the periodic orbit terms for the time delay do not give similar periodic orbit
terms for the conductance.

\section*{Acknowledgements}

The authors would like to thank Eugene Bogomolny, Jon Keating, Sebastian M\"uller,
Alfredo Ozorio de Almeida and Raul Vallejos for helpful discussions and EPSRC for
financial support. M.S.\ wishes to thank the Centro Brasileiro de Pesquisas
F\'isicas for the kind hospitality during a sabbatical leave during which a 
major part of this research was carried out.

\end{document}